# Experimental long-lived entanglement of two macroscopic objects.

Brian Julsgaard, Alexander Kozhekin, and Eugene S. Polzik

*Institute of Physics and Astronomy, University of Aarhus, 8000 Aarhus, Denmark*

**Entanglement is considered to be one of the most profound features of quantum mechanics[1,2]. An entangled state of a system consisting of two subsystems cannot be described as a product of the quantum states of the two subsystems[9,10,16,17]. In this sense the entangled system is considered inseparable and nonlocal. It is generally believed that entanglement manifests itself mostly in systems consisting of a small number of microscopic particles. Here we demonstrate experimentally the entanglement of two objects, each consisting of about $10^{12}$ atoms. Entanglement is generated via interaction of the two objects - more precisely, two gas samples of cesium atoms - with a pulse of light, which performs a non-local Bell measurement on collective spins of the samples[14]. The entangled spin state can be maintained for 0.5 millisecond. Besides being of fundamental interest, the robust, long-lived entanglement of material objects demonstrated here is expected to be useful in quantum information processing, including teleportation[3-5] of quantum states of matter and quantum memory.**

In 1935 Einstein, Podolsky and Rosen (EPR) published a famous paper[1] formulating what they perceived as a paradox created by quantum mechanics. Since then, the EPR correlations and other types of entanglement have been extensively analyzed with seminal contributions made by J. Bell[2]. Entangled or inseparable states are fundamental to the field of quantum information, specifically to quantum teleportation of discrete[3,4] and continuous[5] variables and to quantum dense coding of discrete[6] and continuous[7,8] variables, to name a few examples. The majority of experiments on entanglement to date deal with entangled states of light[3-6,9-11]. Entangled states of discrete photonic variables (spin ½ systems)[9,10] as well as entangled states of continuous variables (quadrature-phase operators) of the electro-magnetic field[11] have been generated experimentally. Entangled states of material particles are much more difficult to generate experimentally; however, such states are vital for storage and processing of quantum information. Recently, entangled states of four trapped ions have been produced[12], and two atoms have been entangled via interaction with a microwave photon field[13].

In this Letter we describe an experiment on the generation of entanglement between two separate samples of atoms containing $10^{12}$ atoms each, along the lines of a recent proposal[14]. Besides the fact that we demonstrate a quantum entanglement at the level of macroscopic objects, our experiment proves feasible a new approach to the quantum interface between light and atoms suggested in[14,15] and paves the road towards the other protocols proposed there, such as the teleportation of atomic states and quantum memory. The entanglement is generated through a non-local Bell measurement on the two samples' spins performed by transmitting a pulse of light through the samples.

The ideal EPR entangled state of two sub-systems described by continuous non-commuting variables $\hat{X}_{1,2}$ and $\hat{P}_{1,2}$, e.g., positions and momenta of two particles, is the state for which $\hat{X}_1 + \hat{X}_2 \to 0, \hat{P}_1 - \hat{P}_2 \to 0$. Recently in[16,17], the necessary and sufficient condition for the entanglement or inseparability for such Gaussian quantum variables has been cast in a form of an inequality involving only the variances of variables:

$\left\langle (\delta(\hat{X}_1 + \hat{X}_2))^2 \right\rangle + \left\langle (\delta(\hat{P}_1 - \hat{P}_2))^2 \right\rangle < 2$. In our experiment the quantum variables analogous to the position and momentum operators are two projections of the collective spin (total angular momentum) of an atomic sample. The analogy is evident from the commutation relation $\left[ \hat{J}_z, \hat{J}_y \right] = i\hat{J}_x$, which can be rewritten as $\left[ \hat{X}, \hat{P} \right] = i$, where $\hat{X} = \hat{J}_z / \sqrt{J_x}$ and $\hat{P} = \hat{J}_y / \sqrt{J_x}$ if the atomic sample is spin-polarized along the $x$ axis with $\hat{J}_x$ having a large classical value $J_x$. For two spin-polarized atomic samples with $J_{x1} = -J_{x2} = J_x$ the above entanglement condition translates into

$$\delta J_{12}^2 \equiv \delta J_{z12}^2 + \delta J_{y12}^2 < 2J_x \qquad (1)$$

where we introduce the notations $\delta J_{z12}^2 = \left\langle \left(\delta(\hat{J}_{z1} + \hat{J}_{z2})\right)^2 \right\rangle, \delta J_{y12}^2 = \left\langle \left(\delta(\hat{J}_{y1} + \hat{J}_{y2})\right)^2 \right\rangle$. The interpretation of condition (1) comes from the recognition of the fact that for both atomic samples in coherent spin states (CSS) the equality $\delta J_{y,z}^2 = \frac{1}{2} J_x$ holds. Entanglement between atoms of the two samples is, therefore, according to (1) equivalent to the spin variances smaller than that for samples in a CSS[18,19] characterized by uncorrelated individual atoms. The entangled state of this type is a two-mode squeezed state for the continuous spin variables[14,16].

Entanglement is produced via interaction of atoms with polarized light. A polarized pulse of light is described by Stokes operators obeying the same commutation relation as spin operators, $\left[ \hat{S}_y, \hat{S}_z \right] = i\hat{S}_x$. $\hat{S}_x$ is the difference between photon numbers in $x$ and $y$ linear polarizations, $\hat{S}_y$ is the difference between polarizations at $\pm 45^0$, and $\hat{S}_z$ is the difference between the left- and right-hand circular polarizations along the propagation direction, $z$. In our experiment light is linearly polarized along the $x$ axis. Hence the two pairs of continuous quantum variables engaged in the entanglement protocol are $\hat{J}_z$ and $\hat{J}_y$ for atoms and $\hat{S}_z$ and $\hat{S}_y$ for light.

In this Letter we report on the generation of a state of two separate cesium gas samples (Fig. 1), which obeys the entanglement condition (1). As shown in[14,15], when an off-resonant pulse is transmitted through two atomic samples with opposite mean spins $J_{x1} = -J_{x2} = J_x$, the light and atomic variables evolve as

$$\hat{S}_y^{out} = \hat{S}_y^{in} + \alpha \hat{J}_{z12}, \hat{S}_z^{out} = \hat{S}_z^{in}$$
$$\hat{J}_{y1}^{out} = \hat{J}_{y1}^{in} + \beta \hat{S}_z^{in}, \hat{J}_{y2}^{out} = \hat{J}_{y2}^{in} - \beta \hat{S}_z^{in}, \hat{J}_{z1}^{out} = \hat{J}_{z1}^{in}, \hat{J}_{z2}^{out} = \hat{J}_{z2}^{in} \qquad (2)$$

The first line describes the Faraday effect (polarization rotation of the probe). The second line shows the back action of light on atoms, i. e., spin rotation due to the angular momentum of light. According to (2), the measurement of $\hat{S}_y^{out}$ reveals the value of $\hat{J}_{z12} = \hat{J}_{z1} + \hat{J}_{z2}$ (provided the constant $\alpha$ is large enough, so that $\hat{S}_y^{in}$ is relatively small) without changing this value. It follows from (2) that the total $y$ spin projection for both samples is also conserved, $\hat{J}_{y1}^{out} + \hat{J}_{y2}^{out} = \hat{J}_{y1}^{in} + \hat{J}_{y2}^{in}$ [14]. The procedure can be repeated with another pulse of light measuring the sum of $y$ components, $\hat{J}_{y1} + \hat{J}_{y2}$, again in a non-demolition way, while at the same time leaving the previously measured value of $\hat{J}_{z1} + \hat{J}_{z2}$ intact. As a result, the sum of the $z$ components and the sum of the $y$ components of spins of the two samples are known exactly in the ideal case, and therefore the two samples are entangled according to (1), since the uncertainties on the left-hand side become negligible.

An important modification of the above protocol is the addition of a magnetic field oriented along the direction $x$, which allows to use a single entangling pulse to measure both $z$ and $y$, as described in the Methods section.

The schematic of the experimental set-up is shown in Fig. 1. The two cells are coated from inside with paraffin coating which enhances the ground state coherence time $T_2$ up to $5-30 m\sec$ depending on the density of atoms. The $|J_x|$, $T_2$ and $p$ are measured by the magneto-optical resonance method [21].

After the samples are prepared in CSS, optical pumping is switched off and a probe pulse is sent through. Its Stokes operator $S_y$ is measured by a polarizing beam splitter with two balanced detectors. The differential photocurrent from the detectors is split in two and its $\cos(\Omega t)$ and $\sin(\Omega t)$ power spectral components $\left(S_{yCos}^{out}(\Omega)\right)^2$ and $\left(S_{ySin}^{out}(\Omega)\right)^2$ are measured by lock-in amplifiers. Repeating this sequence many times we obtain the variances for these components, in short, spectral variances, which according to eq.(3) are

$\left(\delta S_{yCos}^{out}(\Omega)\right)^2 = \left(\delta S_{yCos}^{in}(\Omega)\right)^2 + \frac{1}{2}\alpha^2 \delta \hat{J}_{z12}^2 \equiv \frac{1}{2}\delta S^2 + \kappa \delta J_{z12}^2$ and similarly for $\left(\delta S_{ySin}^{out}(\Omega)\right)^2$

with substitution $J_z \to J_y$. The coefficient $\alpha = \sigma \gamma n / 4 F A \Delta \approx 2.5$ is estimated [15] using $\sigma \approx \lambda^2 / 2\pi$ as the resonant dipole cross section, $\gamma = 5 MHz$ - the full width of the optical transition, $A = 2 cm^2$ - the probe beam cross section, and $n = 10^{13}$ - the number of photons in the $0.45 m\sec$ probe pulse with the power of $5 mW$.

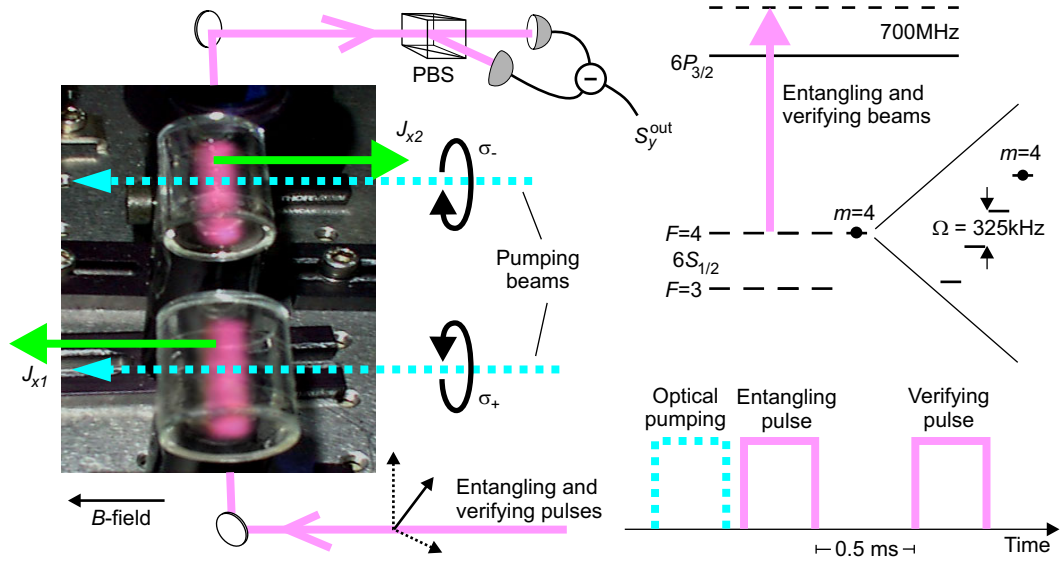

**Figure 1.** Schematic of the experimental setup, atomic level structure and the sequence of optical pulses. Fluorescence of atoms interacting with light is seen inside the cells (artificial colour). The two atomic samples in glass cells 3*3cm at approximately room temperature are placed in a highly homogenous magnetic field of 0.9 Gauss ($\Omega = 325 kHz$) surrounded by a magnetic shield. Cesium atoms are optically pumped into $F = 4, m_F = 4$ ground state in the first cell and into $F = 4, m_F = -4$ in the second cell to form coherent spin states oriented along $x$ axis for cell 1 and along $-x$ for cell 2. Optical pumping is achieved with circularly polarized $0.45 m\sec$ pulses at $852 nm$ and $894 nm$ with opposite helicity for the two cells. The mean spin value for each sample is given by $J_x = N \sum_{m=-4}^{4} m \rho_{mm} = 4Np \to 4N$, with the limiting value corresponding to a perfect spin polarization, $p = 1$. The $|J_x|$ for the two cells is adjusted to be equal to better than 10%. After the optical pumping is completed $0.45 m\sec$ long entangling and verifying pulses separated by $0.5 m\sec$ delay are sent through. Both pulses are blue-detuned by $\Delta = 700 MHz$ from the closest hyperfine component of the $D_2$ line at $852 nm$. The thermal atomic motion is not an obstacle in this experiment but is rather helpful. The probe beam covers most of but not the whole volume of the atomic sample, but the duration of the pulses is longer than the transient time of an atom across the cell. Therefore each light pulse effectively interacts with all atoms of the samples. In addition, a large detuning of the probe makes the Doppler broadening insignificant.

The spectral variance data,
$\Delta = \left(\delta S_{yCos}^{out}(\Omega)\right)^2 + \left(\delta S_{ySin}^{out}(\Omega)\right)^2 = \delta S^2 + \kappa \delta J_{z12}^2 + \kappa \delta J_{y12}^2 \equiv \delta S^2 + \kappa \delta J_{12}^2$, is plotted in Fig. 2. The linear fit to it, $\Delta(J_x)$, is the quantum limit of noise corresponding to the coherent spin state of samples (see figure caption for details).

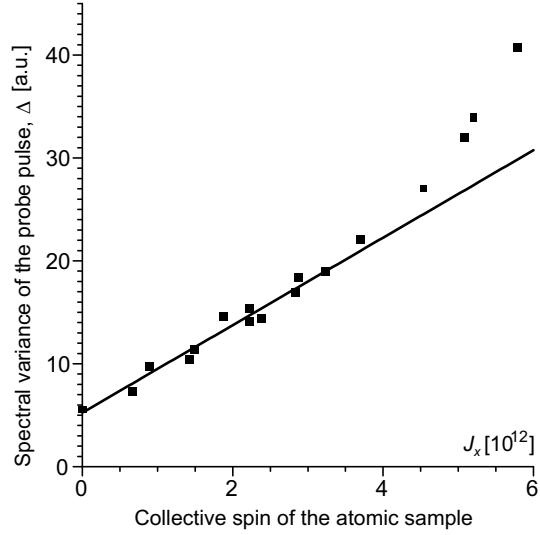

**Figure 2.** Determination of the coherent spin state limit for entanglement. The total measured variance of the two quadratures of the photocurrent
$\Delta = \left(\delta S_{yCos}^{out}(\Omega)\right)^2 + \left(\delta S_{ySin}^{out}(\Omega)\right)^2 = \delta S^2 + \kappa \delta J_{z12}^2 + \kappa \delta J_{y12}^2 \equiv \delta S^2 + \kappa \delta J_{12}^2$ is plotted as a function of $J_x$. $J_x$ is measured independently by magneto-optical resonance method and is varied by heating the cells and by adjusting optical pumping. In the absence of atoms in the cells the measured spectral variance is due to the initial probe state variance $\delta S^2$. $\delta S^2$ is at the vacuum (shot) noise level, which has been verified experimentally by checking its characteristic linear dependence on the probe power. With atoms present the measured variance grows linearly with $J_x$ at low densities, which proves that there is no classical contribution to the spin noise and therefore the observed atomic fluctuations at these densities are entirely due to quantum atomic noise[20]. The degree of the spin polarization for the data in the Figure is nearly perfect, $p \geq 95\%$, which means that the spin state is very close to the coherent spin state (CSS). We therefore conclude that a linear fit to the observed data corresponds to the CSS of both atomic samples for which $\delta J_{12}^2 = 2J_x$ and hence this line, which we denote by $\Delta(J_x)$, establishes the noise level corresponding to the right hand side of the inequality (4). Deviations of the observed variance from the linear fit at higher atomic numbers are due to the nonlinearly growing contribution of classical technical noise of the spin state because of the technical noise of lasers, the non-ideal cancellation of the back action of the probe on the two samples, etc.

Using $\Delta(J_x)$ as a reference level we can now devise a measurement procedure, which will verify the presence of an entangled state. Namely, if for a certain state of the two ensembles the spectral variance of the signal $\Delta_{EPR} = \delta S^2 + \kappa \delta J_{EPR}^2$ obeys the inequality

$$\Delta_{EPR} = \delta S^2 + \kappa \delta J_{EPR}^2 < \delta S^2 + 2\kappa J_x = \Delta(J_x) \qquad (4)$$

then apparently $\delta J_{EPR}^2 < 2J_x$ holds for such a state and therefore this state is entangled in accordance with the condition (1). It is, of course, necessary to use otherwise identical conditions for measurements of $\Delta_{EPR}$ and $\Delta(J_x)$, i.e. the same value of $2J_x$ and the same probe intensity and detuning in order to keep the constant $\kappa$ unchanged throughout the entire experiment. As mentioned above the value of $2J_x$ has been controlled by a magneto-optical resonance measurement to better than 5%. In the actual experiment described below the measurements of $\Delta_{EPR}$ and $\Delta(J_x)$ have been conducted in turn at identical conditions at the repetition rate of $500Hz$.

The measurement sequence aimed at the generation and verification of the entanglement consists of the optical pumping pulses preparing samples in a CSS, the *entangling* pulse preparing the samples in the entangled state (pulse I) and the *verifying* pulse coming after the delay time $\tau$ and verifying the entanglement (pulse II). These pulses have the same duration and optical frequency as the probe pulse used for the CSS measurements. Between the two pulses the joint spin state of the two samples is subject to decoherence. The photocurrents from the two pulses are subtracted electronically and the variance of the difference, $\Delta_{EPR}$, is measured (see the Methods section). The vanishing $\Delta_{EPR}$ corresponds to two repeated measurements on the total spin state of the two samples producing the same results, i.e. to a perfect knowledge of both $\hat{J}_{z12}$ and $\hat{J}_{y12}$ and therefore to a perfectly entangled state. In the experiment the minimal value of $\Delta_{EPR}$ is $2\delta S^2$ due to the entangling and verifying pulses quantum noise.

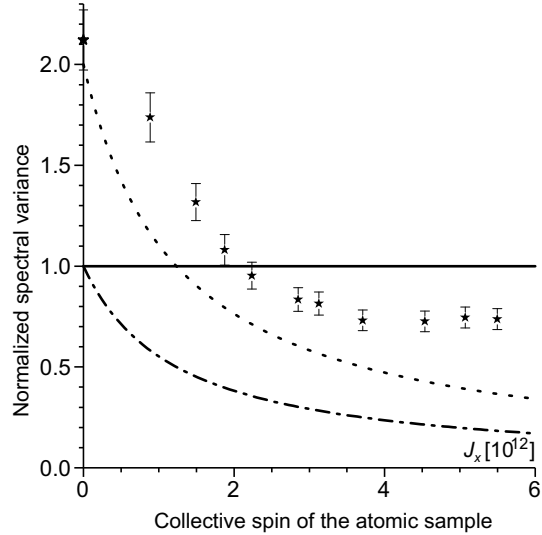

**Figure 3.** Demonstration of the entangled spin state for two atomic samples. The results are plotted as a function of $J_x$ and are normalized to the CSS limit $\Delta(J_x)$ (the linear fit in Fig. 2). This limit - the boundary between entangled (below the line) and separable states - thus corresponds to the unity level in the figure (solid line). The raw experimental data $\Delta_{EPR}/\Delta(J_x)$ – data for the entangled spin state, which has lived for $\tau = 0.5m\sec$ - are shown as stars. The values below the unity level verify that the entangled state of the two atomic samples has been generated and maintained for $0.5m\sec$. The minimal possible level for $\Delta_{EPR}/\Delta(J_x)$ (maximum possible entanglement) is equal to $2\delta S^2/\Delta(J_x)$ (dashed line), i.e. it is set by the total quantum noise of the two pulses. The normalized shot noise level of the verifying pulse, $\delta S^2/\Delta(J_x)$ (dashed-dotted line), which is used for calculations of the degree of entanglement is also shown.

The results of measurements with $\tau = 0.5 m\sec$ are shown in Fig. 3. The results are normalized to the CSS limit $\Delta(J_x)$ (the linear fit in Fig. 2). This limit thus corresponds to the unity level in Fig. 3. The raw experimental data $\Delta_{EPR}/\Delta(J_x)$ for the entangled state are shown as stars. The values below the unity level verify that the entangled state of the two atomic samples has been generated and maintained for $0.5 m\sec$. The derivation of the degree of entanglement from the data in Fig. 3 is given in the Methods section. The degree of entanglement calculated operationally from the data without additional assumptions is $\xi = (35 \pm 7)\%$. The degree of entanglement useful for teleportation calculated using an additional, experimentally proven assumption of the initially CSS for both samples is higher, $\xi_{\exp er} = (52 \pm 7)\%$. The predicted teleportation fidelity performed with the same two pulses as used in the present paper (see Methods section) is $F = 55\%$, which is above the classical limit of $50\%$. Factors limiting the fidelity are of technical nature, and we expect higher fidelity to be well within reach.

The imperfect entanglement comes from several factors. Firstly, the vacuum noise of the entangling pulse prohibits a perfect preparation of the spin state especially for a small number of atoms. Secondly, the deviation of the initial spin state from CSS, which is due to the classical noise of the lasers, becomes more pronounced as the number of atoms grows. Finally, losses of light on the way from one cell to another, as well as the spin state decay between the two measurements due to decoherence caused by quadratic Zeeman splitting and collisions preclude perfect entanglement at all atomic densities.

Measurements with delay times longer than $0.8 m\sec$ demonstrate no entanglement. The decoherence process can be illustrated by visualizing two squeezed ellipses in the phase space, corresponding to the two samples, with the size along anti-squeezed axes of the order of 3 in units of the coherent state (experimentally measured value). When the ellipses start to dephase (rotate) the total noise along the squeezed axis becomes equal to the coherent state noise (zero degree of entanglement) after the time of the order of 1.2 msec if the initial entanglement of 65% (maximal degree allowed by the experimental light noise) is assumed. After 0.6 msec delay the degree of entanglement of 35% should be expected. These numbers agree reasonably well with observations.

It is instructive to analyze the difference between the degree of entanglement and the degree of classical correlations. For uncorrelated atomic samples the normalized variance of the difference between the two photocurrents would be equal to 2 in notations of Fig.3. The pure atomic part of this variance $V_{uncorr}$ would be approximately 1.5 for medium atomic densities. For the actual data in Fig. 3 the atomic part of the variance is approximately $V_{corr} = 0.25$. The degree of correlation, $1 - V_{corr}/V_{uncorr}$, is 83% for this example. This number is much higher than the degree of entanglement implying that entanglement requires something stronger than classical correlations. The reason for this difference is that in quantum mechanics the measuring device (light in this case) becomes entangled with the measured object and therefore the noise of these two subsystems cannot be treated independently.

In conclusion, we have demonstrated on-demand generation of entanglement of two separate macroscopic objects, which can be maintained for more than $0.5 m\sec$. The state we have demonstrated is not a maximally entangled "Schrödinger cat" state, which for $10^{12}$ atoms would not survive even for a femtosecond under conditions of our experiment. Our state is similar to a two mode squeezed state, and is an example of a non-maximally entangled state which is suitable for a particular purpose, e.g., for the atomic teleportation. The long life time of this multi-particle entanglement is due to a high symmetry of the generated state. Entanglement manifests itself only in the collective properties of the two ensembles. Therefore a loss of coherence for a single atom makes a negligible effect on the entanglement, unlike, e.g., in case of maximally entangled multi-particle state. The entanglement is generated by means of light propagating through the two samples and therefore the samples can be rather distant, as required for communications. The off-resonant character of the interaction used for the creation of entanglement allows for the potential extension of this method to other media, possibly including solid state samples with long lived spin states.

## Methods.

**Entangling measurement of spin components in the rotating frame with a single light pulse.** The Larmor precession of the $J_z, J_y$ components with a common frequency $\Omega$ does not change their mutual orientation and size and therefore does not affect the entanglement. On the other hand, the precession allows us to extract information about both $z$ and $y$ components from a single probe pulse as described below. Moreover, in the lab frame the spin state is now encoded at the frequency $\Omega$, and as usual an ac measurement is easier to reduce to the quantum noise level than a dc measurement. Measurements of the light noise can be now conducted only around its $\Omega$ spectral component. By choosing a suitable radio-frequency value for $\Omega$ we can reduce the probe noise $\hat{S}_{y,z}^{in}(\Omega)$ to the minimal level of the vacuum (shot) noise. In the presence of the magnetic field the spin behavior is described by the following equations: $\dot{\hat{J}}_z(t) = \Omega \hat{J}_y(t), \dot{\hat{J}}_y(t) = -\Omega \hat{J}_z(t) + \beta \hat{S}_z(t)$, whereas the Stokes operators still evolve according to equation (2). Solving the spin equations and using (2) we obtain

$$\hat{S}_y^{out}(t) = \hat{S}_y^{in}(t) + \alpha \left\{ \hat{J}_{z12} \cos(\Omega t) + \hat{J}_{y12} \sin(\Omega t) \right\} \tag{3}$$

The $\hat{J}_{z,y}$ components are now defined in the frame rotating with the frequency $\Omega$ around the magnetic field direction $x$. It is clear from (3) that by measuring the $\cos(\Omega t)/\sin(\Omega t)$ component of $\hat{S}_y^{out}(t)$ one acquires the knowledge of the $z/y$ spin projections. Simultaneous measurement of both spin components of the two atomic samples is possible because they commute, $[\hat{J}_{z1} + \hat{J}_{z2}, \hat{J}_{y1} + \hat{J}_{y2}] = i(J_{x1} + J_{x2}) = 0$.

**Degree of entanglement.** The differential noise for the entangling (pulse I) and verifying (pulse II) pulses can be expressed using (3) as

$$\Delta_{EPR} = \delta \bar{S}_{Cos}^2 + \delta \bar{S}_{Sin}^2 \equiv \left( \delta S_{yCos}^{Iout}(\Omega) - \delta S_{yCos}^{IIout}(\Omega) \right)^2 + \left( \delta S_{ySin}^{Iout}(\Omega) - \delta S_{ySin}^{IIout}(\Omega) \right)^2 =$$
$$= \delta S_{II}^2 + \kappa \left\{ \kappa^{-1} \delta S_I^2 + \left( \delta(\hat{J}_{z12})_I - \delta(\hat{J}_{z12})_{II} \right)^2 + \left( \delta(\hat{J}_{y12})_I - \delta(\hat{J}_{y12})_{II} \right)^2 \right\} \equiv \tag{5}$$
$$\equiv \delta S_{II}^2 + \kappa \delta J_{EPR}^2$$

The variance $\delta J_{EPR}^2$ determines the uncertainty of the spin components of the state prepared by the entangling pulse and measured after the time $\tau$. It contains two contributions, $\delta S_I^2$ due to the quantum noise of the entangling pulse, and all other terms in curly brackets, which are due to decoherence during the time $\tau$. The verifying pulse in our experiment is also not noiseless and therefore its variance, $\delta S_{II}^2$, also contributes to $\Delta_{EPR}$. In the experiment $\delta S_I^2 = \delta S_{II}^2 = \delta S^2$. The exact entanglement condition, $\Delta_{EPR} < \Delta(J_x)$, is obtained by substituting (5) into (4). The degree of entanglement can be defined as $\xi = 1 - \frac{\delta J_{EPR}^2}{2J_x} = 1 - \frac{\Delta_{EPR} - \delta S^2}{\Delta(J_x) - \delta S^2}$. $\xi$ varies from 0 (separable state) to 1 (perfect entanglement). The highest degree of entanglement calculated operationally from the data is $\xi = (35 \pm 7)\%$.

An alternative calculation of the degree of entanglement can be carried out based on ref. [14], which takes into account that the initial state of both samples is characterized by the CSS noise level. Eq. (2) of [14] in notations of the present paper yields for the highest possible degree of entanglement defined again as $\xi = 1 - \frac{\delta J_{EPR}^2}{2J_x}$ and created by the first pulse in our protocol, the value $\xi_{theory} = 1 - \eta_{theory}$ where the degree of the "two mode squeezing" is, according to [14], $\eta_{theory} = \frac{\delta S^2}{\delta S^2 + 2\kappa J_x} = \frac{\delta S^2}{\Delta(J_x)}$. This is the best possible entanglement for a given ratio of the light noise and the atomic spin noise contribution. This value of entanglement would correspond to the minimal possible difference between the two pulses, $\Delta_{EPR}^{min} = 2\delta S^2$. Note that this value is closer to unity (stronger entanglement) than the degree of entanglement $\xi = 1 - \frac{\Delta_{EPR}^{min} - \delta S^2}{\Delta(J_x) - \delta S^2} = 1 - \frac{\delta S^2}{\Delta(J_x) - \delta S^2}$ calculated according to the operational definition used in the previous paragraph. The reason for that is that the entanglement in [14] is calculated assuming that the samples prior to entanglement are known to be in the CSS. Since this fact has been experimentally verified in the present paper for up to intermediate atomic densities, we may use this way of calculating the degree of entanglement at these densities as well. However, we are interested in the degree of entanglement, which has survived the delay time and which has been measured by the verifying pulse. Due to various reasons, including the decoherence during the delay time, the differential noise between the two pulses does not reduce to its minimal value, $\Delta_{EPR} > \Delta_{EPR}^{min} = 2\delta S^2$. Therefore the actual degree of entanglement, as witnessed by the measurement, is lower than maximal possible, $\xi_{exper} = 1 - \eta_{exper} < \xi_{theory}$. Here the degree of squeezing, $\eta_{exper} = \frac{\Delta_{EPR} - \delta S^2}{\Delta(J_x)} > \eta_{theory}$ obtained from the data in Fig. 3 is worse than the theoretical best value, $\eta_{theory}$, due in part to the diffusion of the state during the delay time between the entangling and the verifying pulses. From Fig. 3 we find $\eta_{exper} = 0.48$ and therefore $\xi_{exper} = 52\%$ for $J_x \approx 3.5 \times 10^{12}$.

**Expected fidelity of teleportation.** It is also possible to estimate the fidelity of teleportation which would be achieved if the second, verifying pulse would instead of verification be used for the Bell measurement on one of the entangled samples and a sample to-be-teleported. Using the value of $\xi_{\exp er}$ and the equation (3) from [14] we obtain for the fidelity of teleportation $F = 55\%$ for $J_x \approx 3.5 \times 10^{12}$. This is higher than the classical boundary of $F = 50\%$.

Note that the entangled state reported here has a random element in it, namely, every (ideal) entangling measurement creates the state $J_{z1} + J_{z2} = x, J_{y1} + J_{y2} = p$ with $x, p$ being random measured values. However, this state is as efficient for, e.g., the teleportation as is a state $J_{z1} + J_{z2} = 0, J_{y1} + J_{y2} = 0$.

We gratefully acknowledge the contributions of J. Hald, J. L. Sørensen, C. Schori and A. Verchovski. We are also grateful to I. Cirac, A. Kuzmich, A. Sørensen, and P. Zoller for illuminating discussions.